\documentclass[aps,pra,showpacs,twocolumn]{revtex4-1}
\usepackage{amssymb}
\usepackage{amsmath}
\usepackage{graphicx}
\usepackage{epsfig}

\begin{document}

\title{Majorana charges, winding numbers and Chern numbers in quantum Ising
models}
\author{G. Zhang${}^1$${}^2$, C. Li${}^1$ and Z. Song${}^1$}
\email{songtc@nankai.edu.cn}
\affiliation{${}^1$School of Physics, Nankai University, Tianjin 300071, China \\
${}^2$College of Physics and Materials Science, Tianjin Normal University,
Tianjin 300387, China}

\begin{abstract}
Mapping a many-body state on a loop in parameter space is a simple way to
characterize a quantum state. The connections of such a geometrical
representation to the concepts of Chern number and Majorana zero mode are
investigated based on a generalized quantum spin system with short and
long-range interactions. We show that the topological invariants, the Chern
numbers of corresponding Bloch band is equivalent to the winding number in
the auxiliary plane, which can be utilized to characterize the phase
diagram. We introduce the concept of Majorana charge, the magnitude of which
is defined by the distribution of Majorana fermion probability in zero-mode
states, and the sign is defined by the type of Majorana fermion. By direct
calculations of the Majorana modes we analytically and numerically verify
that the Majorana charge is equal to Chern numbers and winding numbers.
\end{abstract}

\maketitle



\section{Introduction}

Characterizing the quantum phase transitions (QPTs) is of central
significance to both condensed matter physics and quantum information
science. Exactly solvable quantum many-body models are benefit to
demonstrate the concept and characteristic of QPTs. Recently, topological
phases and phase transitions \cite{Wen} have attracted much attention in
various physical contexts. In general, QPTs are classified two types,
characterized by topologically nontrivial properties in the Hilbert space,
and by the local order parameters associated with symmetry breaking \cite%
{Sachdev}, respectively. Both conventional and topological QPTs refer to\
the sudden change of the groundstate properties driven by the change of
external parameters. A\ topological QPT involves the change of ground-state
topological properties which are indicated by topological quantum discrete
numbers \cite{Kane,Zhang}, while the various phases in a conventional QPT
are distinguished by continuously varying order parameters. The topological
quantum number is topological invariant, such as Chern number and\ Majorana
zero mode, which\ have been received much recent interest \cite%
{Wilczek,Moore,Nayak,Ivanov,kitaev,Fu,Alicea,Oreg,Roy,Sen,Niu,Chen1,Chen2}.

Nevertheless, so far there are no evidences to suggest that the two types of
QPTs are absolutely incompatible, not occur at the same point for certain
systems. An interesting question is whether the local order parameter and
topological order parameter can coexist to characterize the quantum phase
transitions. In recent work \cite{Gang}, it is shown that the variation of
the groundstate energy density for a class of exactly solvable quantum Ising
models, which is a function of a loop in a two-dimensional auxiliary space,
experiences a nonanalytical point when the winding number of the
corresponding loop changes. This fact indicates that this class of models
can be joint ones in which a topological and a conventional QPTs occur
simultaneously.

In this paper, we investigate topological properties in a family of exactly
solvable Ising models with short- and long-range interactions. We introduce
the concept of Majorana charge to indicate the phase diagram based on the
corresponding Majorana tight-binding lattice with open boundary condition.
The magnitude of Majorana charge is determined by the distribution of
Majorana fermion probability in zero-mode states, while its sign is
determined by the types of Majorana fermions.\ We show that the topological
invariants, the Chern numbers of a corresponding Bloch band equal to the
winding number in the auxiliary plane. Furthermore by direct calculations of
the Majorana modes we analytically and numerically verify that the Majorana
charge is equal to Chern numbers and winding numbers. These indicate that
three quantities can equally characterize the phase diagram in quantum spin
systems.

This paper is organized as follows. In section. \ref{Model and pseudo-spin
representation}, we present a generalized one-dimensional quantum spin
model, which is exactly solvable by introducing a pseudo-spin
representation. In section. \ref{Chern and winding numbers}, we show that
the Chern number and winding number are identical. Section. \ref{Majorana
charge of zero mode} investigates the Majorana fermion representation of the
models. Section. \ref{Summary} summarizes the results and explores its
implications.

\begin{center}
\begin{figure*}[tbp]
\includegraphics[ bb=14 111 458 479, width=0.32\textwidth, clip]{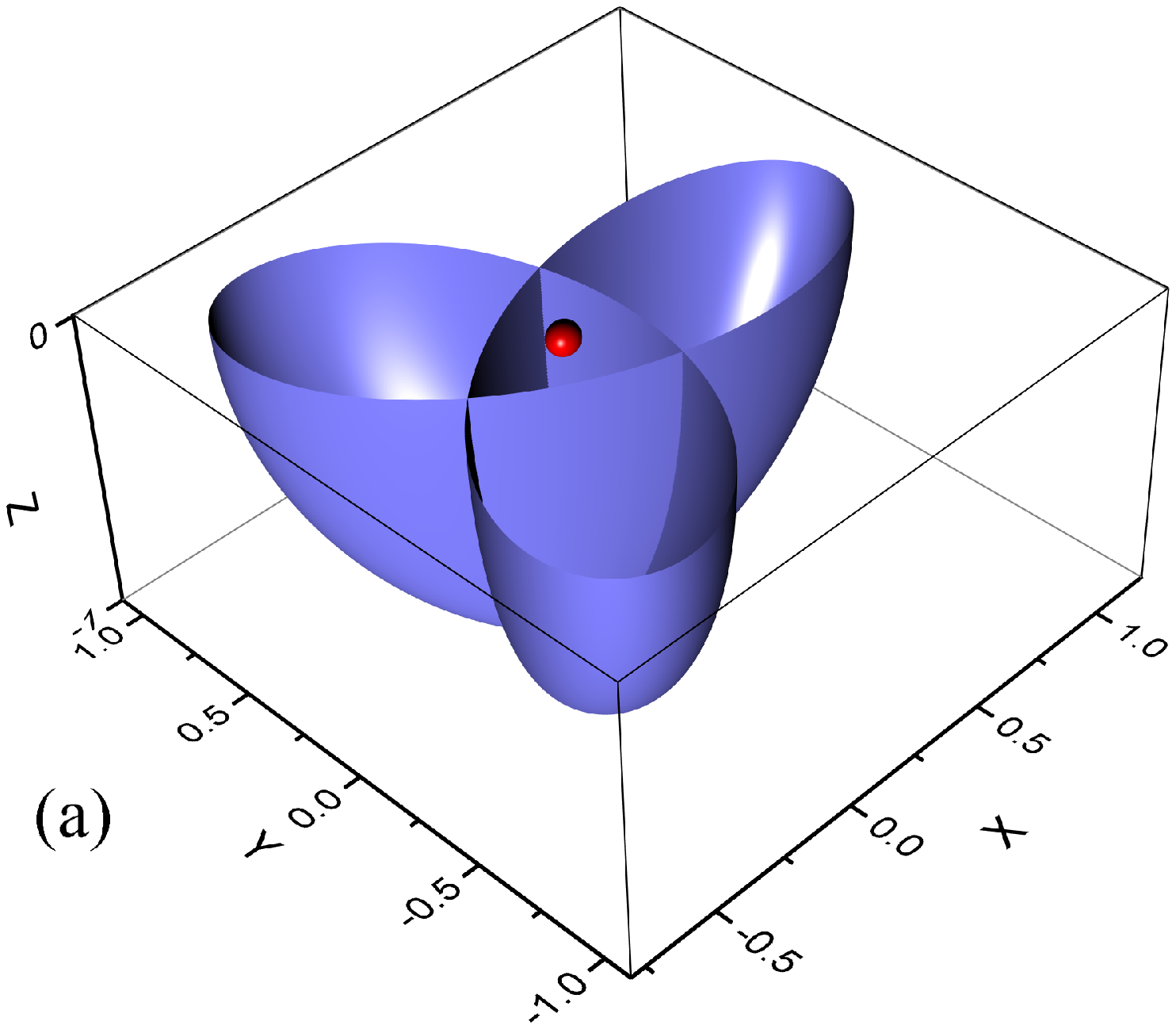} %
\includegraphics[ bb=14 111 458 479, width=0.32\textwidth, clip]{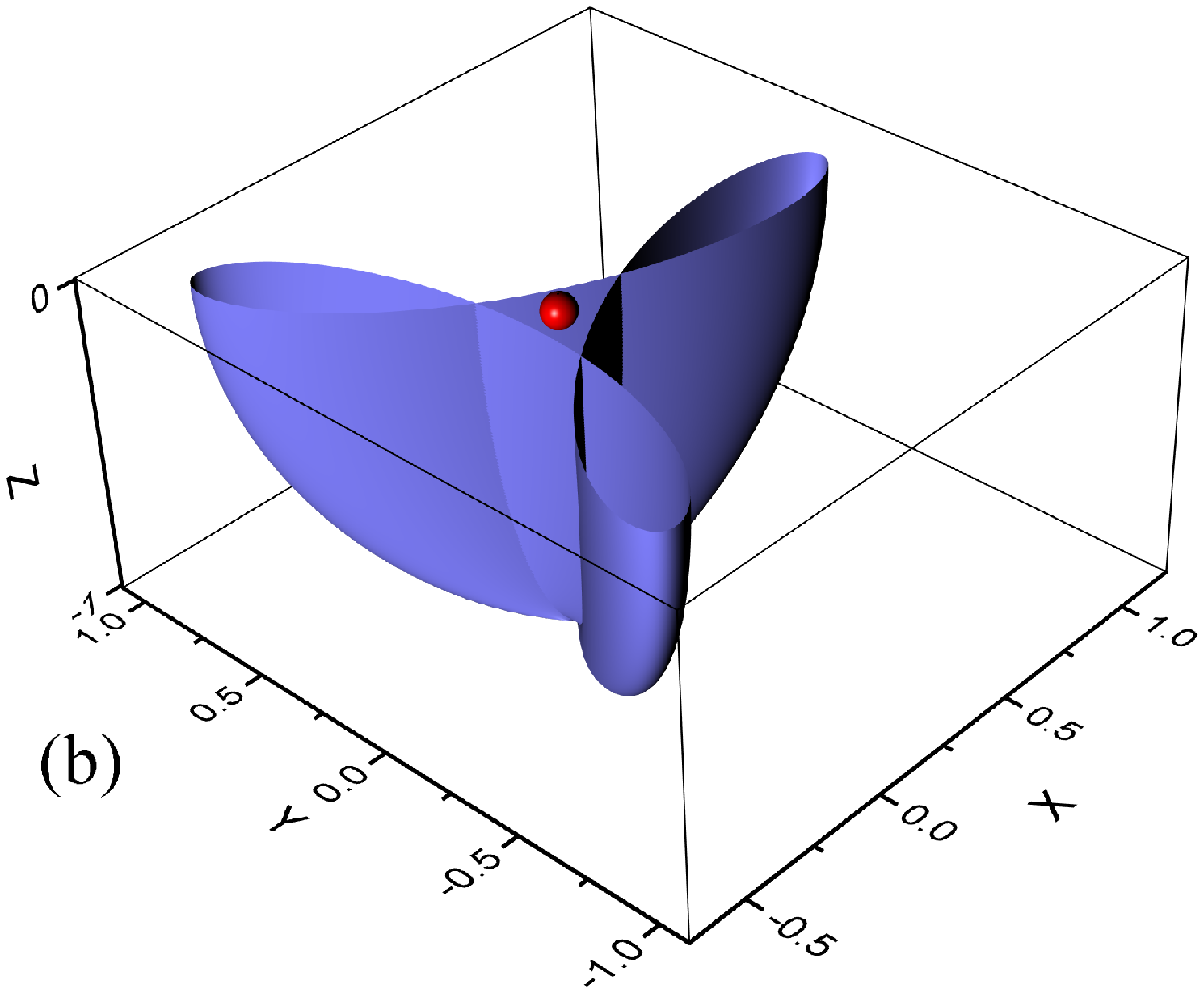} %
\includegraphics[ bb=14 111 458 479, width=0.32\textwidth, clip]{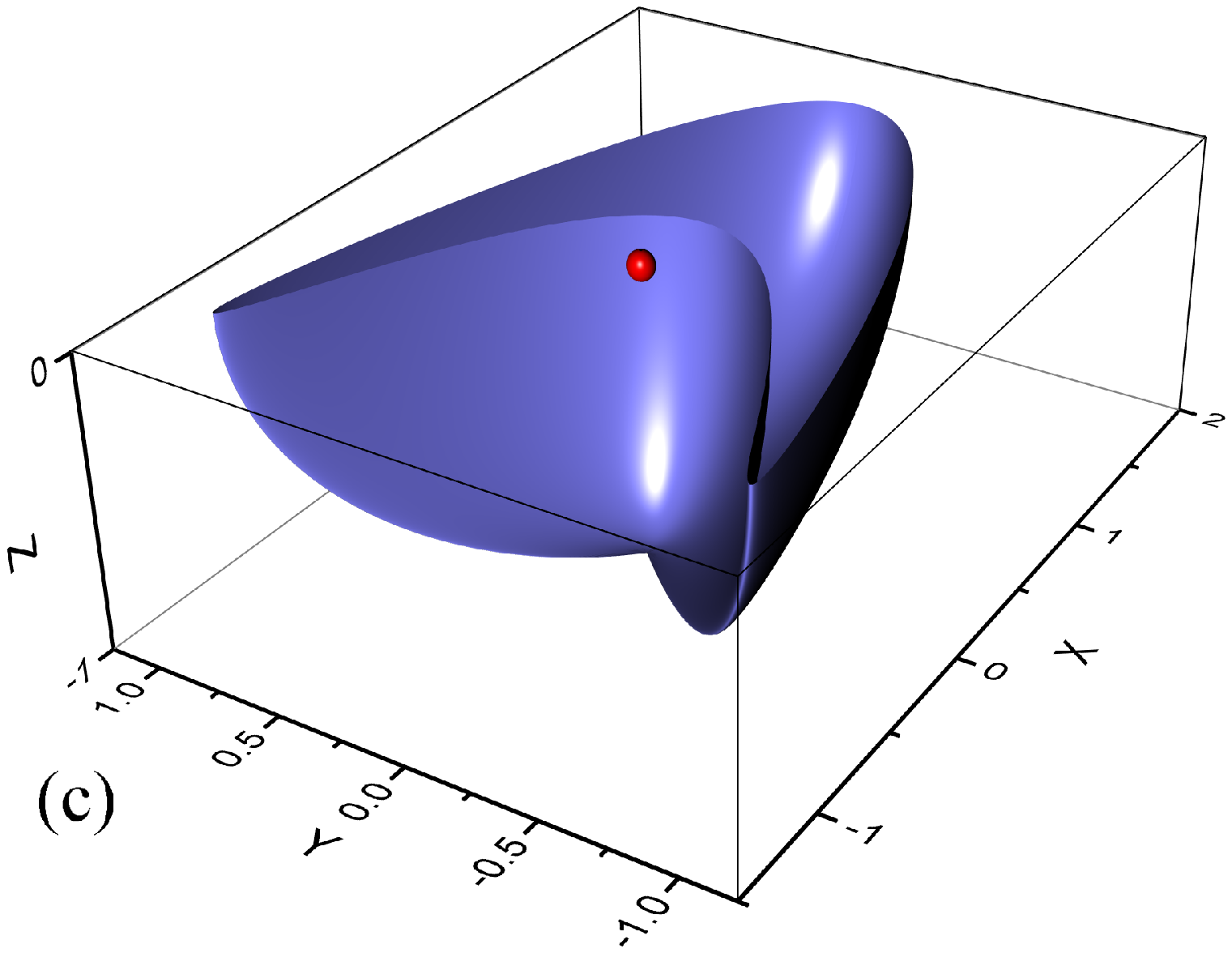} %
\includegraphics[ bb=27 137 458 452, width=0.32\textwidth, clip]{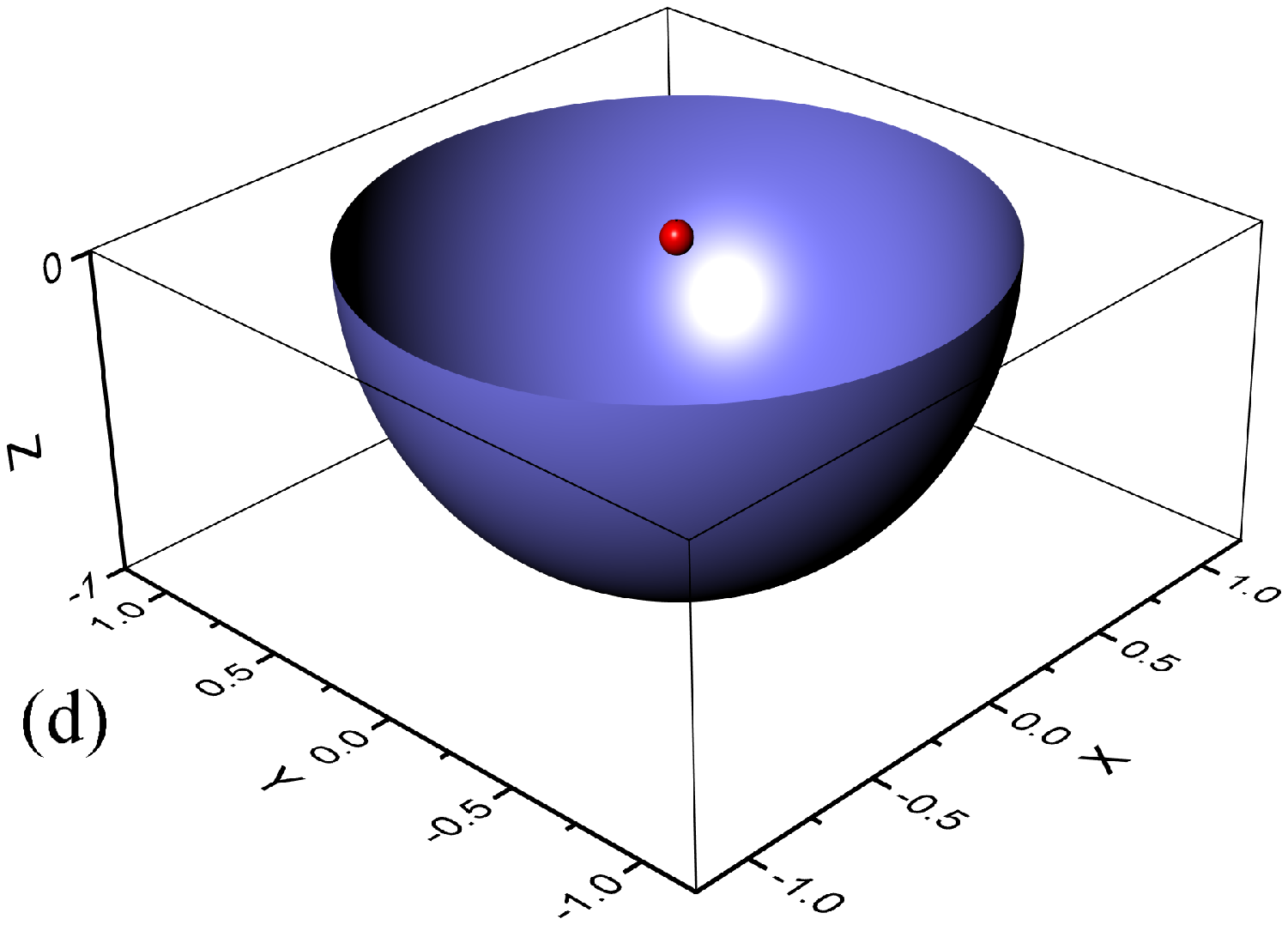} %
\includegraphics[ bb=27 137 458 452, width=0.32\textwidth, clip]{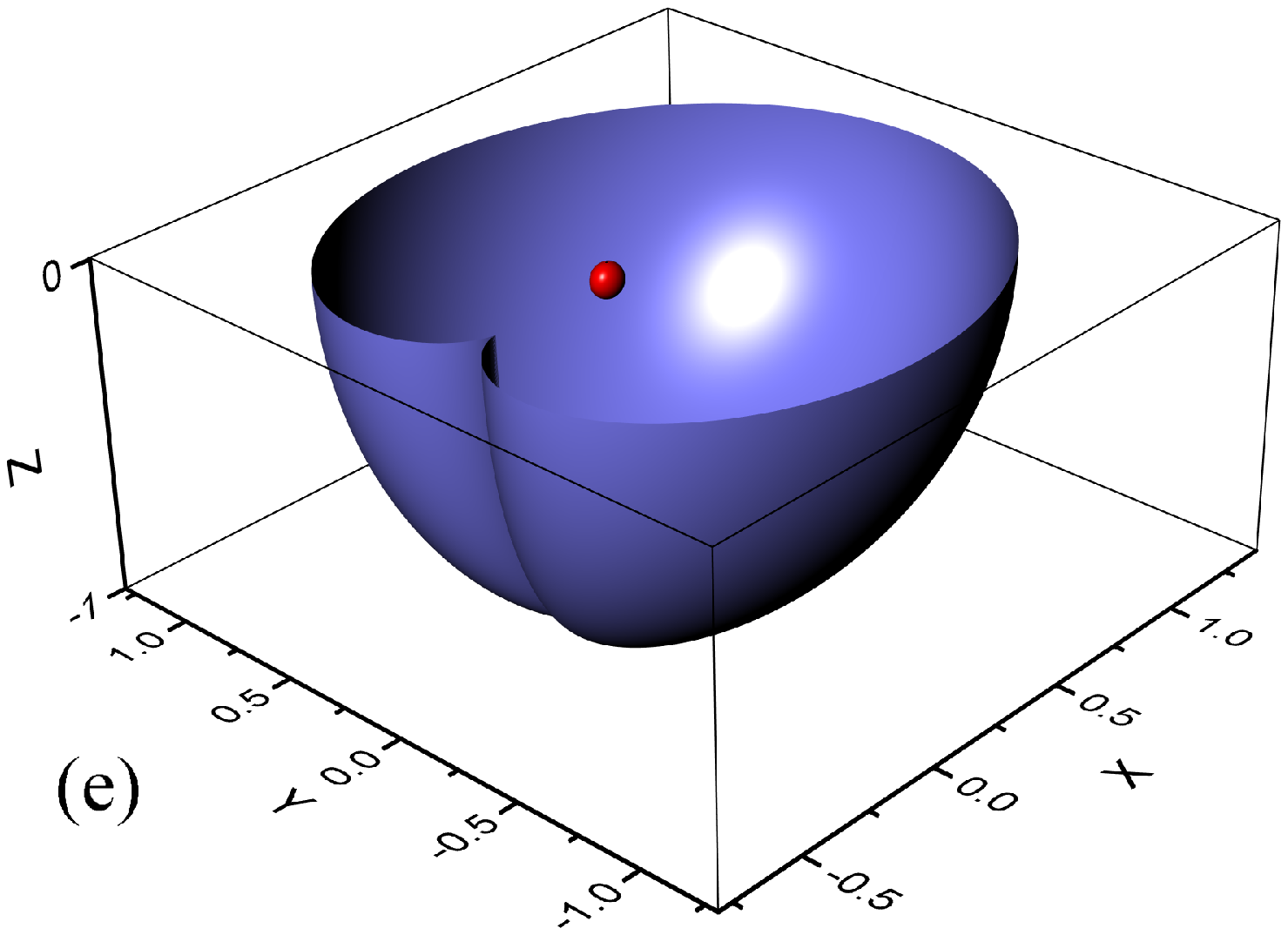} %
\includegraphics[ bb=27 137 458 452, width=0.32\textwidth, clip]{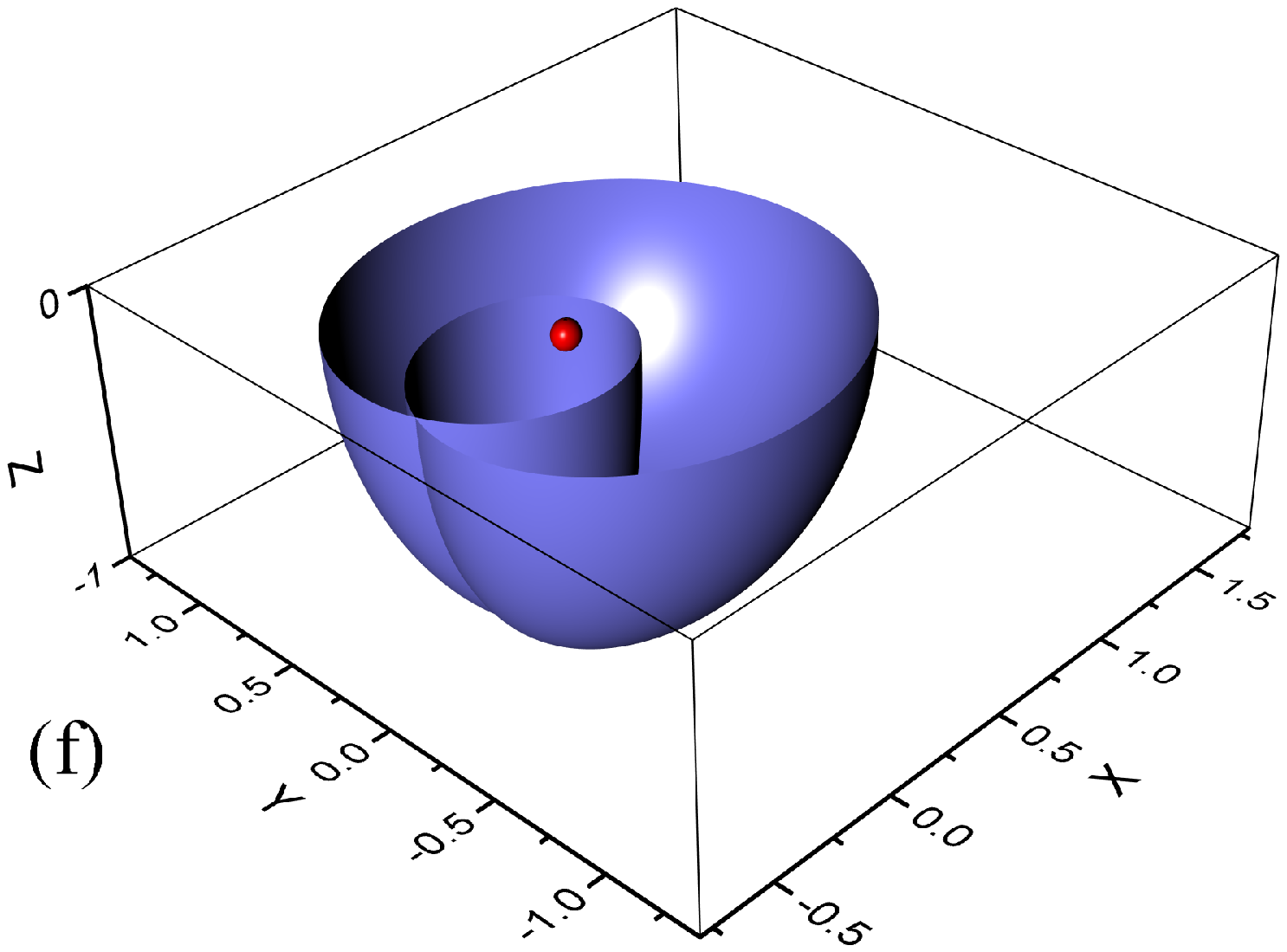} %
\includegraphics[ bb=28 95 455 488, width=0.32\textwidth, clip]{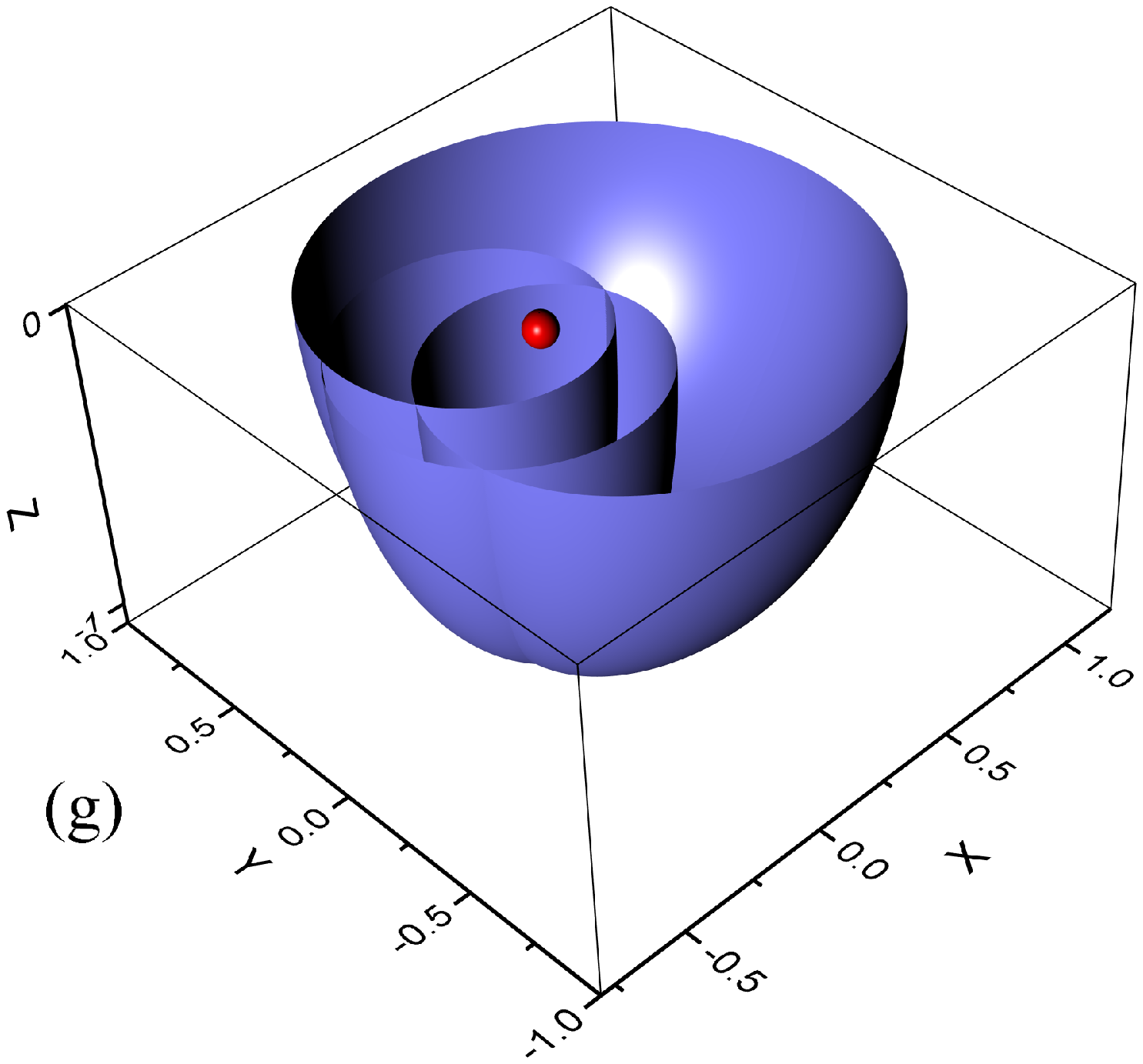} %
\includegraphics[ bb=28 95 455 488, width=0.32\textwidth, clip]{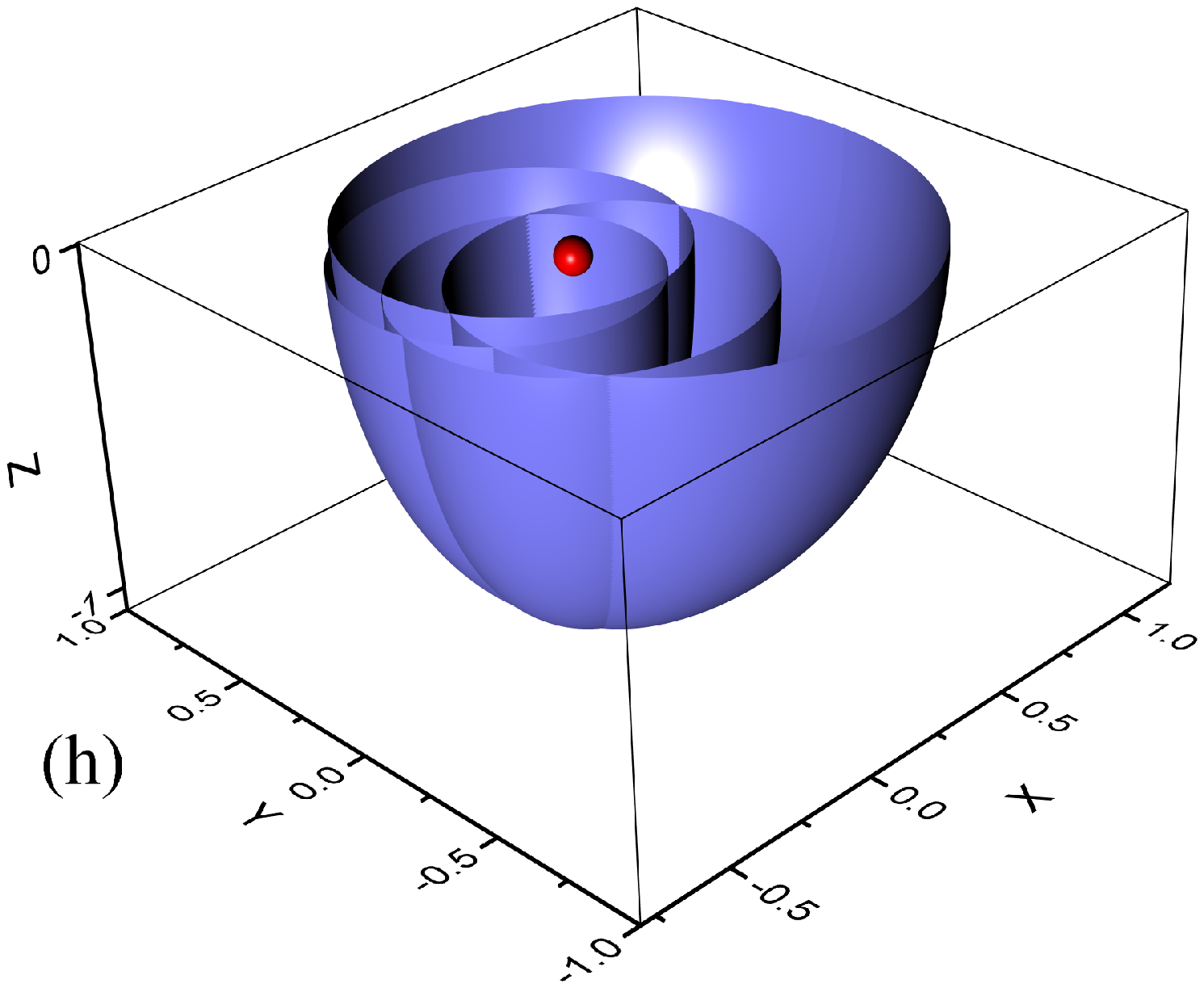} %
\includegraphics[ bb=28 95 455 488, width=0.32\textwidth, clip]{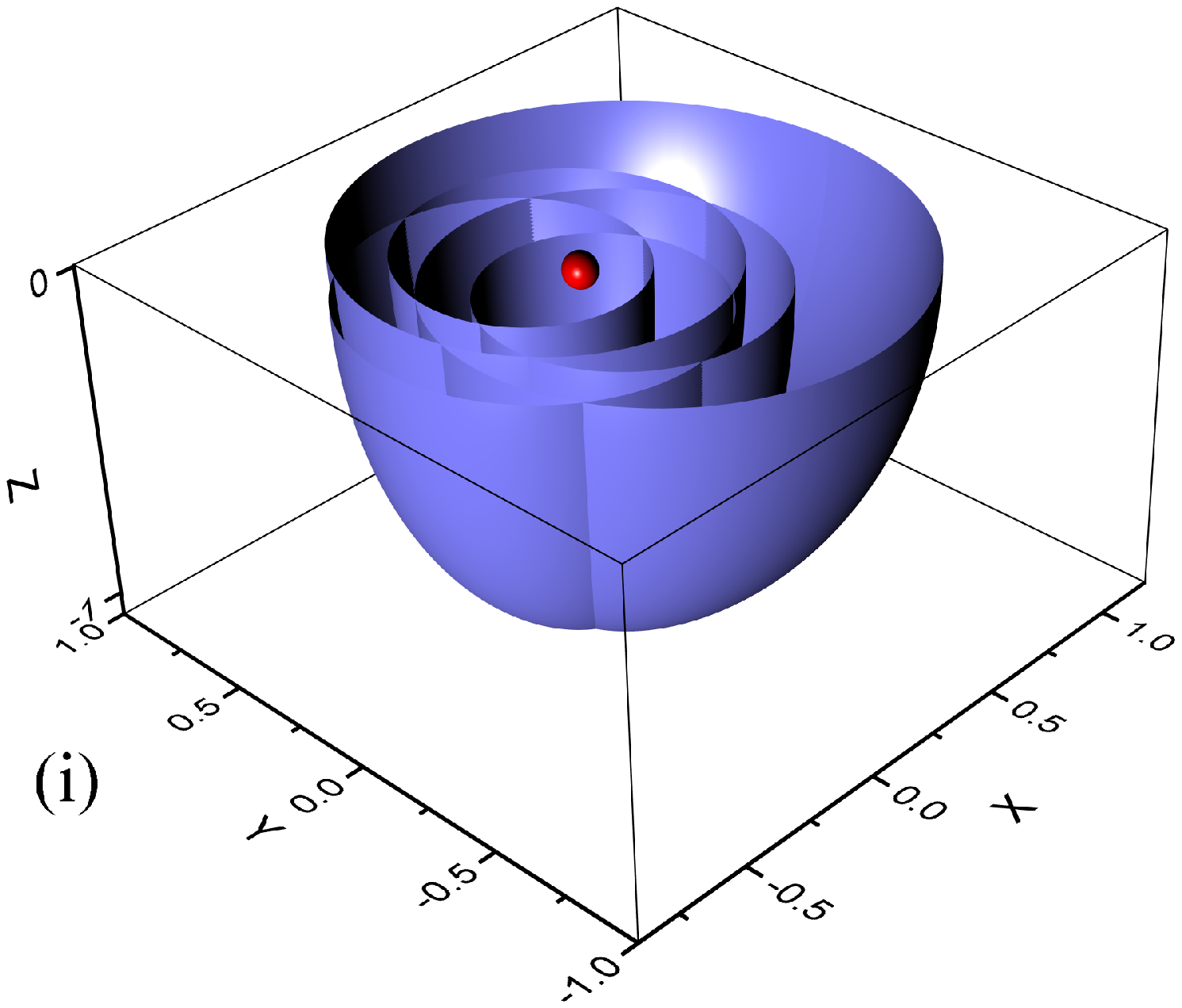}
\caption{(Color online) Plots of the surfaces in the auxiliary space ($x,y,z$%
)\ to illustrate the relation between winding and Chern number. The
corresponding parameters in equations of the 3D surfaces are listed in Table
I. The red dot denotes the origin of the auxiliary space ($0,0,0$). The
winding numbers can be figured out from the curves in the $xy$ plane.}
\label{fig1}
\end{figure*}
\end{center}

\begin{table*}[tbp]
\caption{Typical examples illustrating the relations among winding numbers,
zero modes and Majoraana charges. The values of $J_{n}^{x}$ and $J_{n}^{y}$\
($n\in \lbrack 1,$ $5]$) are the parameters for equations of plots in Fig.
\protect\ref{fig1} with figure index (FI) (a)-(i) and corresponding numbers $%
\mathcal{N}$. For finite size systems with $N=200$ and $500$, the zero modes
and Majorana charges are obtained by exact diagonalizations. We define the
zero modes by selecting eigenstates with absolute eigenvalues less than $%
10^{-10}$.\ $N_{\mathrm{zm}}$ is the number of such eigenstates for every
cases. The Majorana charges are calculated from Eq. (\protect\ref{charge})
for given zero mode states. We can see that $N_{\mathrm{zm}}=2\left\vert
\mathcal{N}\right\vert $ and $\mathcal{M}$\ closes to $2\mathcal{N}$\ as $N$
increases.}
\label{Table I}
\begin{center}
\renewcommand\arraystretch{1}
\par
\begin{tabular}{ccccccccc}
\hline\hline
FI\  & $\ \ \mathcal{N}$ \  & \ $J_{1}^{x},$ $J_{1}^{y}$ \ \  & \ $%
J_{2}^{x}, $ $J_{2}^{y}$ \ \  & $\ \ J_{3}^{x},$ $J_{3}^{y}$ \ \  & $\ \
J_{4}^{x},$ $J_{4}^{y}$ \ \  & $\ \ J_{5}^{x},$ $J_{5}^{y}$ \ \  & $N_{%
\mathrm{zm}}$ & $\mathcal{M}$ $(N=200,$ $500)$ \\ \hline
a & $-2$ & $0.4,$ $0$ & $0,$ $0.6$ & $0,$ $0$ & $0,$ $0$ & $0,$ $0$ & $4$ & $%
-3.65,$ $-3.86$ \\
b & $-1$ & $0,$ $0.55$ & $0.45,$ $0$ & $0,$ $0$ & $0,$ $0$ & $0,$ $0$ & $2$
& $-1.71,$ $-1.88$ \\
c & $0$ & $0.8,$ $-0.2$ & $0.5,$ $0.5$ & $0,$ $0$ & $0,$ $0$ & $0,$ $0$ & $0$
& $0$ \\
d & $1$ & $1,$ $0$ & $0,$ $0$ & $0,$ $0$ & $0,$ $0$ & $0,$ $0$ & $2$ & $2,$ $%
2$ \\
e & $1$ & $0.8,$ $0$ & $0.4,$ $0$ & $0,$ $0$ & $0,$ $0$ & $0,$ $0$ & $2$ & $%
2,$ $2$ \\
f & $2$ & $0.4,$ $0$ & $0.6,$ $0$ & $0,$ $0$ & $0,$ $0$ & $0,$ $0$ & $4$ & $%
3.92,$ $3.9678$ \\
g & $3$ & $0.3,$ $0$ & $0.2,$ $0$ & $0.5,$ $0$ & $0,$ $0$ & $0,$ $0$ & $6$ &
$5.76,$ $5.9033$ \\
h & $4$ & $0.25,$ $0$ & $0.1,$ $0$ & $0.15,$ $0$ & $0.5,$ $0$ & $0,$ $0$ & $%
8 $ & $7.70,$ $7.88$ \\
i & $5$ & $0.2,$ $0$ & $0,$ $0$ & $0.15,$ $0$ & $0.15,$ $0$ & $0.5,$ $0$ & $%
10$ & $9.38,$ $9.77$ \\ \hline\hline
\end{tabular}%
\end{center}
\end{table*}

\section{Model and pseudo-spin representation}

\label{Model and pseudo-spin representation}We consider a generalized
one-dimensional quantum spin model, which was exactly solved four decades
ago \cite{Suzuki}. It contains long-range interactions and the Hamiltonian
has the form
\begin{eqnarray}
H &=&\sum\limits_{n=1}^{M}\sum\limits_{j=1}^{N}\left( J_{n}^{x}\sigma
_{j}^{x}\sigma _{j+n}^{x}+J_{n}^{y}\sigma _{j}^{y}\sigma _{j+n}^{y}\right)
\notag \\
&&\times \prod_{l=j+1}^{j+n-1}\sigma _{l}^{z}+g\sum\limits_{j=1}^{N}\sigma
_{j}^{z}  \label{H}
\end{eqnarray}%
The operators $\sigma _{i}^{x,y,z}$ are the Pauli matrices for spin at $i$th
site. In the case $M=1$, it is reduced to an ordinary anisotropic $XY$
model, which has been employed as a platform to test the signatures of QPT,
such as entanglement \cite{Fazio}, geometric phase \cite{Pachos,Zhu},
decoherence \cite{Quan}, and fidelity \cite{Zanardi}.\ In large $N$ limit, $%
M\ll N$, the Hamiltonian can be diagonalized as the form

\begin{equation}
H=\sum_{k}\epsilon _{k}\left( \gamma _{k}^{\dagger }\gamma _{k}-\frac{1}{2}%
\right) ,
\end{equation}%
via a conventional Jordan-Wigner transformation
\begin{eqnarray}
\sigma _{j}^{z} &=&1-2c_{j}^{\dagger }c_{j}\text{, }\sigma _{j}^{y}=\mathrm{i%
}\sigma _{j}^{x}\sigma _{j}^{z},  \label{JW1} \\
\sigma _{j}^{x} &=&-\prod\limits_{l<j}\left( 1-2c_{l}^{\dagger }c_{l}\right)
\left( c_{j}+c_{j}^{\dag }\right) ,  \label{JW2}
\end{eqnarray}%
and Fourier transformation%
\begin{equation}
c_{j}=\frac{1}{\sqrt{N}}\sum_{k}c_{k}e^{ikj},  \label{Fourier}
\end{equation}%
and a Bogoliubov transformation%
\begin{equation}
c_{k}=u_{k}\gamma _{k}+iv_{k}\gamma _{-k}^{\dagger }.  \label{B T}
\end{equation}%
Here $\gamma _{k}$ is a fermion operator and the parameters are%
\begin{equation}
u_{k}=\cos \frac{\theta _{k}}{2},v_{k}=\sin \frac{\theta _{k}}{2},
\label{UV}
\end{equation}%
with%
\begin{eqnarray}
\cos \theta _{k} &=&\frac{2}{\epsilon _{k}}[g-\sum\limits_{n=1}^{M}\left(
J_{n}^{x}+J_{n}^{y}\right) \cos \left( nk\right) ], \\
\sin \theta _{k} &=&\frac{2}{\epsilon _{k}}\sum\limits_{n=1}^{M}\left(
J_{n}^{x}-J_{n}^{y}\right) \sin \left( nk\right) .
\end{eqnarray}%
The spectrum is in the form
\begin{eqnarray}
\epsilon _{k} &=&2\{[\sum\limits_{n=1}^{M}\left( J_{n}^{x}-J_{n}^{y}\right)
\sin \left( nk\right) ]^{2}  \notag \\
&&+[\sum\limits_{n=1}^{M}\left( J_{n}^{x}+J_{n}^{y}\right) \cos \left(
nk\right) -g]^{2}\}^{1/2},  \label{spectrum}
\end{eqnarray}%
where $k\in \lbrack -\pi ,\pi )$. Based on this analysis, the groundstate
phase diagram can be obtained. Actually, for $k=k_{c}=0,$\ we have%
\begin{equation}
\epsilon _{k_{c}}=2\left\vert \sum\limits_{n=1}^{M}\left(
J_{n}^{x}+J_{n}^{y}\right) -g\right\vert .
\end{equation}%
And for $k=k_{c}=\pi ,$\ we have%
\begin{equation}
\epsilon _{k_{c}}=2\left\vert \sum\limits_{n=1}^{M}\left( -1\right)
^{n}\left( J_{n}^{x}+J_{n}^{y}\right) -g\right\vert .
\end{equation}%
We find that the derivatives of $\epsilon _{k_{c}}$\ with respect to
parameters $\{J_{n}^{x},J_{n}^{y},g\}$\ experience a discontinuity at points%
\begin{equation}
g=g_{c}=\sum\limits_{n=1}^{M}\left( J_{n}^{x}+J_{n}^{y}\right) .  \label{g_c}
\end{equation}%
or%
\begin{equation}
g=g_{c}=\sum\limits_{n=1}^{M}\left( -1\right) ^{n}\left(
J_{n}^{x}+J_{n}^{y}\right) .
\end{equation}

In this paper, we will consider the phase diagram in alternative ways:
pseudo spin and Majorana fermion representations. This starting point is the
spinless fermion Hamiltonian%
\begin{equation}
H=H_{\mathrm{ch}}+H_{\mathrm{b}},
\end{equation}%
with%
\begin{eqnarray}
H_{\mathrm{ch}} &=&\sum\limits_{n=1}^{M}\sum\limits_{j=1}^{N-n}[\left(
J_{n}^{x}+J_{n}^{y}\right) c_{j}^{\dag }c_{j+n}+\left(
J_{n}^{x}-J_{n}^{y}\right) c_{j}^{\dag }c_{j+n}^{\dag }  \notag \\
&&+\mathrm{H.c.]}+\sum\limits_{j=1}^{N}\left( g-2gc_{j}^{\dagger
}c_{j}\right) ,  \label{Hch}
\end{eqnarray}%
and%
\begin{eqnarray}
H_{\mathrm{b}} &=&\left( -1\right)
^{N_{p}+1}\sum\limits_{n=1}^{M}\sum\limits_{j=N-n+1}^{N}[\left(
J_{n}^{x}+J_{n}^{y}\right) c_{j}^{\dag }c_{j+n}  \notag \\
&&+\left( J_{n}^{x}-J_{n}^{y}\right) c_{j}^{\dag }c_{j+n}^{\dag }]+\mathrm{%
H.c.},  \label{Hbc}
\end{eqnarray}%
which represent the chain and the boundary parts, respectively. Here, $%
N_{p}=\sum\limits_{j=1}^{N}c_{j}^{\dagger }c_{j}$ is the number of fermion.

In large $N$ limit and $M\ll N$ , the Hamiltonian can be written in $k$
space as

\begin{eqnarray}
&&H=\sum\limits_{k}\{2[\sum\limits_{n=1}^{M}\left(
J_{n}^{x}+J_{n}^{y}\right) \cos \left( nk\right) -g]c_{k}^{\dag }c_{k}
\notag \\
&&-\mathrm{i}\sum\limits_{n=1}^{M}\left( J_{n}^{x}-J_{n}^{y}\right) \sin
\left( nk\right) \left( c_{-k}^{\dag }c_{k}^{\dag }+c_{-k}c_{k}\right) +g\}
\end{eqnarray}%
\bigskip by performing the Jordan-Wigner transformation in Eq. (\ref{JW1},%
\ref{JW2}) and Fourier transformation in Eq. (\ref{Fourier}), respectively.\
We notice that the spinless fermion Hamiltonian is actually an extended
one-dimensional mean field model for a triplet superconductor with
long-range hopping.

An alternative way to diagonalize the Hamiltonian $H$\ is to induce the
pseudo spin
\begin{eqnarray}
s_{k}^{-} &=&\left( s_{k}^{+}\right) ^{\dag }=c_{k}c_{-k},  \notag \\
s_{k}^{x} &=&\frac{1}{2}\left( c_{k}^{\dag }c_{k}+c_{-k}^{\dag
}c_{-k}-1\right) ,  \label{pseudospin} \\
s_{k}^{z} &=&\frac{1}{2}\left( s_{k}^{+}+s_{k}^{-}\right) ,  \notag \\
s_{k}^{y} &=&\frac{1}{2i}\left( s_{k}^{+}-s_{k}^{-}\right) ,  \notag
\end{eqnarray}%
instead of Bogoliubov operator\ $\gamma _{k}$.\ These operators satisfy the
commutation relations of Lie algebra%
\begin{equation}
\left[ s_{k}^{x},s_{k^{\prime }}^{\pm }\right] =\pm \delta _{kk^{\prime
}}s_{k^{\prime }}^{\pm },\text{ }\left[ s_{k}^{+},s_{k^{\prime }}^{-}\right]
=2\delta _{kk^{\prime }}s_{k^{\prime }}^{x},
\end{equation}%
and lead to an alternative expression of the Hamiltonian%
\begin{equation}
H=\sum\limits_{k>0}H_{k}=4\sum\limits_{k>0}\overrightarrow{B}\left( k\right)
\cdot \overrightarrow{s}_{k},  \label{H_BS}
\end{equation}%
where the components of $\overrightarrow{B}\left( k\right) $\ are%
\begin{eqnarray}
B_{x} &=&\sum\limits_{n=1}^{M}\left( J_{n}^{x}+J_{n}^{y}\right) \cos \left(
nk\right) -g, \\
B_{y} &=&\sum\limits_{n=1}^{M}\left( J_{n}^{x}-J_{n}^{y}\right) \sin \left(
nk\right) , \\
B_{z} &=&0.
\end{eqnarray}%
The spin of the operator $\overrightarrow{s}_{k}$\ can be taken as $s_{k}=0$%
, $0$,\ and $\frac{1}{2}$. In this paper, we focus on the ground state,
which corresponds to the case with $s_{k}=\frac{1}{2}$\ for all $k$. In this
sense, the physics of the Hamiltonian is clear, which represents an ensemble
of spin-$\frac{1}{2}$ particles in the field of a magnetic monopole. We note
that%
\begin{equation}
\left[ H_{k},H_{k^{\prime }}\right] =0,
\end{equation}%
which indicates that $H_{k}$ is equivalent to the Hamiltonian of two Bloch
bands \cite{Di}.

In recent work \cite{Gang}, it has been generally shown that a system as the
form of Eq. (\ref{H_BS}) can be regarded as an ensemble of free spins on a
loop subjected to a 2D magnetic field of Dirac monopole \cite{Dirac}. The
variation of the groundstate energy density, which is a function of the
loop, experiences a nonanalytical point when the winding number of the
corresponding loop changes. This fact indicates the relation between quantum
phase transition and the geometrical order parameter characterizing the
phase diagram.

The concept of Berry phase can be introduced since $H_{k}$\ can be regarded
as a parameter dependent Hamiltonian. Furthermore, when we consider the band
under a slowly varying time-dependent perturbation, a quantized Berry phase
should be obtained and may characterize the features of the band. As we
shall see, the simplified Hamiltonian provides a natural platform to
investigate the topological characterization of the QPT.

In the following, we consider two Hamiltonians $H_{\mathrm{ch}}+H_{\mathrm{b}%
}$\ and $H_{\mathrm{ch}}$, the ring Hamiltonian and the chain Hamiltonian.
For the ring Hamiltonian, as mentioned above, the translational symmetry
results in $H_{k}$. This ensures the calculations of Chern and winding
numbers, which are utilized to identify the quantum phase. For the chain
Hamiltonian, we will transform $H_{\mathrm{ch}}$\ into Majorana fermion
representation. The phase diagram will be indicated by the number of zero
modes.

\section{Chern and winding numbers}

\label{Chern and winding numbers}We note that the ring Hamiltonian $H_{k}$\
always connects a loop in an auxiliary space. In previous paper \cite{Gang},
it has been shown that when the loop crosses the origin of the auxiliary
space, phase transitions occur. Meanwhile, the winding number of the loop
changes. Then the phase diagram can be characterized by the winding number
of the loop. The conclusion is applicable to the present generalized model,
which corresponds to a loop tracing with the parametric equation $%
\overrightarrow{r}\left( k\right) =\left( x\left( k\right) ,y\left( k\right)
,0\right) $ with\
\begin{equation}
\left\{
\begin{array}{c}
x\left( k\right) =\sum\limits_{n=1}^{M}\left( J_{n}^{x}+J_{n}^{y}\right)
\cos \left( nk\right) -g \\
y\left( k\right) =\sum\limits_{n=1}^{M}\left( J_{n}^{x}-J_{n}^{y}\right)
\sin \left( nk\right)%
\end{array}%
\right. .  \label{pe}
\end{equation}%
The winding number of a closed curve in the auxiliary $xy$-plane around the
origin is defined as

\begin{equation}
\mathcal{N}=\frac{1}{2\pi }\int\nolimits_{c}\frac{1}{r^{2}}\left( x\mathrm{d}%
y-y\mathrm{d}x\right) ,  \label{winding number}
\end{equation}%
which is an integer, representing the total number of times that the curve
travels anticlockwise around the origin. Then we establish the connection
between the QPT and the switch of the topological quantity. Here we present
a class of simple models to illustrate the idea.

We consider a class of Hamiltonian indexed by $n$,
\begin{eqnarray}
H_{n} &=&\sum\limits_{j=1}^{N}\left( J_{n}^{x}\sigma _{j}^{x}\sigma
_{j+n}^{x}+J_{n}^{y}\sigma _{j}^{y}\sigma _{j+n}^{y}\right)  \notag \\
&&\times \prod_{l=j+1}^{j+n-1}\sigma _{l}^{z}+g\sum\limits_{j=1}^{N}\sigma
_{j}^{z},  \label{H_n}
\end{eqnarray}%
which corresponds to a loop tracing with the parametric equation\
\begin{equation}
\left\{
\begin{array}{c}
x_{n}\left( k\right) =\left( J_{n}^{x}+J_{n}^{y}\right) \cos \left(
nk\right) -g \\
y_{n}\left( k\right) =\left( J_{n}^{x}-J_{n}^{y}\right) \sin \left( nk\right)%
\end{array}%
\right. .
\end{equation}%
The geometry of the curve is obvious, which is the superposition of $n$
identical ellipses with winding number $\mathcal{N=}n$ ($\mathcal{-}n$)
according to the Eq. (\ref{winding number}) for $\left\vert g\right\vert
<\left\vert J_{n}^{x}+J_{n}^{y}\right\vert $ and $J_{n}^{x2}-J_{n}^{y2}>0$ ($%
J_{n}^{x2}-J_{n}^{y2}<0$). Similarly, when we consider a Hamiltonian as $%
H_{n}$\ by switching $J_{n}^{x}$ and $J_{n}^{y}$ (or switching $\sigma
_{j}^{x}$\ and $\sigma _{j}^{y}$), the corresponding loop obeys the equation%
\begin{equation}
\left\{
\begin{array}{c}
x_{n}\left( k\right) =\left( J_{n}^{x}+J_{n}^{y}\right) \cos \left(
nk\right) -g \\
y_{n}\left( k\right) =-\left( J_{n}^{x}-J_{n}^{y}\right) \sin \left(
nk\right)%
\end{array}%
\right. ,
\end{equation}%
which still represents $n$ identical ellipses but with winding number $%
\mathcal{N}=-n$\ for $\left\vert g\right\vert <\left\vert
J_{n}^{x}+J_{n}^{y}\right\vert $\textbf{.}

More explicitly,\textbf{\ }when taking\textbf{\ }$J_{n}^{y}=J_{n\neq
1}^{x}=0 $\textbf{\ }and\textbf{\ }$J_{1}^{x}=J^{x}\neq 0$\textbf{, }the
system reduces to ordinary transverse field Ising model with Hamiltonians%
\begin{equation}
H_{\text{\textrm{Ising}}}=\sum\limits_{j=1}^{N}J^{x}\sigma _{j}^{x}\sigma
_{j+n}^{x}+g\sum\limits_{j=1}^{N}\sigma _{j}^{z}.  \label{Ising}
\end{equation}%
The winding number for the ground states of $H_{\text{\textrm{Ising}}}$\
with $\left\vert g\right\vert <\left\vert J^{x}\right\vert $\ is $1$.
Similarly, when taking\textbf{\ }$J_{n}^{x}=J_{n\neq 1}^{y}=0$\textbf{\ }and%
\textbf{\ }$J_{1}^{y}=J^{y}\neq 0$,\textbf{\ }the system reduces to%
\begin{equation}
H_{\text{\textrm{Ising}}}^{\prime }=\sum\limits_{j=1}^{N}J^{y}\sigma
_{j}^{y}\sigma _{j+n}^{y}+g\sum\limits_{j=1}^{N}\sigma _{j}^{z}.
\end{equation}%
The winding number for the ground states of $H_{\text{\textrm{Ising}}%
}^{\prime }$\ with $\left\vert g\right\vert <\left\vert J^{y}\right\vert $\
is $-1$. The opposite signs representing two different quantum phases. These
encouraging results strongly motivate further study of the relation between
quantum phase and the geometric quantity of the system in the auxiliary
space. To this end, we parameterize $\overrightarrow{r}\left( k\right) $ by
its polar angle $\theta $ and azimuthal angle $\varphi $%
\begin{equation}
\overrightarrow{r}\left( k,\varphi \right) =\left( r\sin \varphi \cos \theta
,r\sin \varphi \sin \theta ,\cos \varphi \right) ,  \label{extension1}
\end{equation}%
where $r=\left\vert \overrightarrow{r}\left( k\right) \right\vert $ $=\sqrt{%
x^{2}\left( k\right) +y^{2}\left( k\right) }$and%
\begin{equation}
\sin \theta =\frac{y\left( k\right) }{r},\cos \theta =\frac{x\left( k\right)
}{r}.
\end{equation}%
It is a 2D-to-3D extension for the original model. The corresponding
Hamiltonian can be expressed as%
\begin{equation}
H_{k}\left( \varphi \right) =4\overrightarrow{r}\left( k,\varphi \right)
\cdot \overrightarrow{s}_{k},
\end{equation}%
which goes back to the original one when $\varphi =\pi /2$. The two
eigenstates $\left\vert u_{k}^{\pm }\right\rangle $, with energies $\pm
E=\pm \sqrt{\cos ^{2}\varphi +r^{2}\sin ^{2}\varphi }$, are%
\begin{equation}
\left\vert u_{k}^{\pm }\right\rangle =\frac{1}{\sqrt{2E\left( E\pm \cos
\varphi \right) }}\left(
\begin{array}{c}
\cos \varphi \pm E \\
re^{i\theta }\sin \varphi%
\end{array}%
\right) .
\end{equation}%
We are interested in the ground state, then considering the lower energy
level. The Berry connection is given by%
\begin{eqnarray}
A_{k} &=&i\left\langle u_{k}^{-}\right\vert \partial _{k}\left\vert
u_{k}^{-}\right\rangle  \notag \\
&=&-\frac{1}{2E\left( E-\cos \varphi \right) }r^{2}\sin ^{2}\varphi \frac{%
\partial \theta }{\partial k}, \\
A_{\varphi } &=&i\left\langle u_{k}^{-}\right\vert \partial _{\varphi
}\left\vert u_{k}^{-}\right\rangle =0,
\end{eqnarray}%
and the Berry curvature is%
\begin{equation}
\Omega _{k\varphi }=\partial _{k}A_{\varphi }-\partial _{\varphi }A_{k}=-%
\frac{1}{2E^{3}}r^{2}\sin \varphi \frac{\partial \theta }{\partial k}.
\end{equation}

The corresponding Chern number is%
\begin{eqnarray}
c &=&\frac{1}{2\pi }\int_{0}^{\pi }\text{d}\varphi \int_{0}^{2\pi }\text{d}%
k\Omega _{k\varphi }  \notag \\
&=&-\frac{1}{4\pi }\int_{0}^{2\pi }r^{2}\frac{\partial \theta }{\partial k}%
\text{d}k\int_{-1}^{1}\left( r^{2}-\left( r^{2}-1\right) t^{2}\right) ^{-3/2}%
\text{d}t  \notag \\
&=&-\frac{1}{2\pi }\left[ \theta \left( 2\pi \right) -\theta \left( 0\right) %
\right] ,  \label{chern_number}
\end{eqnarray}%
where $t=\cos \varphi $. It can be seen that the loop of a Hamiltonian is
the intersection of the integral surface on the $xy$\ plane. Then we have
the conclusion%
\begin{equation}
\left\vert c\right\vert =\left\vert \mathcal{N}\right\vert .  \label{c=N}
\end{equation}%
Here we only take the equation for absolute values. This is because that the
extension in Eq. (\ref{extension1}) is not unique. There are many other ways
of 2D-to-3D extension, which can obtain the same result of Eq. (\ref{c=N}).
For example, one can take the extension by

\begin{equation}
\overrightarrow{r}\left( k,\varphi \right) =\left\vert \overrightarrow{r}%
\left( k\right) \right\vert \left( \sin \varphi \cos \theta ,\sin \varphi
\sin \theta ,\cos \varphi \right) ,  \label{extension2}
\end{equation}%
which leads to $c=-N$.\ However, after taking the transformation by
replacement $\theta \rightarrow -\theta $\ or $\varphi \rightarrow -\varphi $%
, we have $c=N$.\ Then the relation between the signs of $c$ and $N$ depends
on the way of the 2D-to-3D extension. Actually, the absolute sign of $c$\ or
$N$\ is meaningless, while the relative sign of them is physical, opposite
signs representing different states. The similar thing\ will happen in the
Majorana charge of zero mode in next section.

Finally, we would like to point that this conclusion is true for any models
in the form $H_{k}\varpropto xs_{x}^{k}+ys_{y}^{k}$.\ For the model in Eq. (%
\ref{H}), the corresponding parameter equations of the integral surfaces are%
\begin{equation}
\left\{
\begin{array}{c}
x=r\sin \varphi \cos \theta , \\
y=r\sin \varphi \sin \theta , \\
z=\cos \varphi ,%
\end{array}%
\right.  \label{surf eq}
\end{equation}%
where polar angle $\theta $ and radius $r$ are explicitly expressed as%
\begin{eqnarray}
&&\tan \theta =\frac{\sum\limits_{n=1}^{M}\left( J_{n}^{x}-J_{n}^{y}\right)
\sin \left( nk\right) }{\sum\limits_{n=1}^{M}\left(
J_{n}^{x}+J_{n}^{y}\right) \cos \left( nk\right) -g}, \\
&&r=\{[\sum\limits_{n=1}^{M}\left( J_{n}^{x}-J_{n}^{y}\right) \sin \left(
nk\right) ]^{2}  \notag \\
&&+[\sum\limits_{n=1}^{M}\left( J_{n}^{x}+J_{n}^{y}\right) \cos \left(
nk\right) -g]^{2}\}^{1/2}.
\end{eqnarray}%
In this paper, we illustrate our conclusion by several typical cases with
the values of $J_{n}^{x}$ and $J_{n}^{y}$\ ($n\in \lbrack 1,5]$) listed in
Table I and plot the 3D surfaces in Fig. \ref{fig1}. The plots display the
relation between the magnitudes of winding and Chern number clearly.
However, the signs of the numbers cannot be visualized in the plots. This
can be done by tracing the plots for varying $k$.

\begin{figure*}[tbp]
\includegraphics[ bb=20 355 575 810, width=0.45\textwidth, clip]{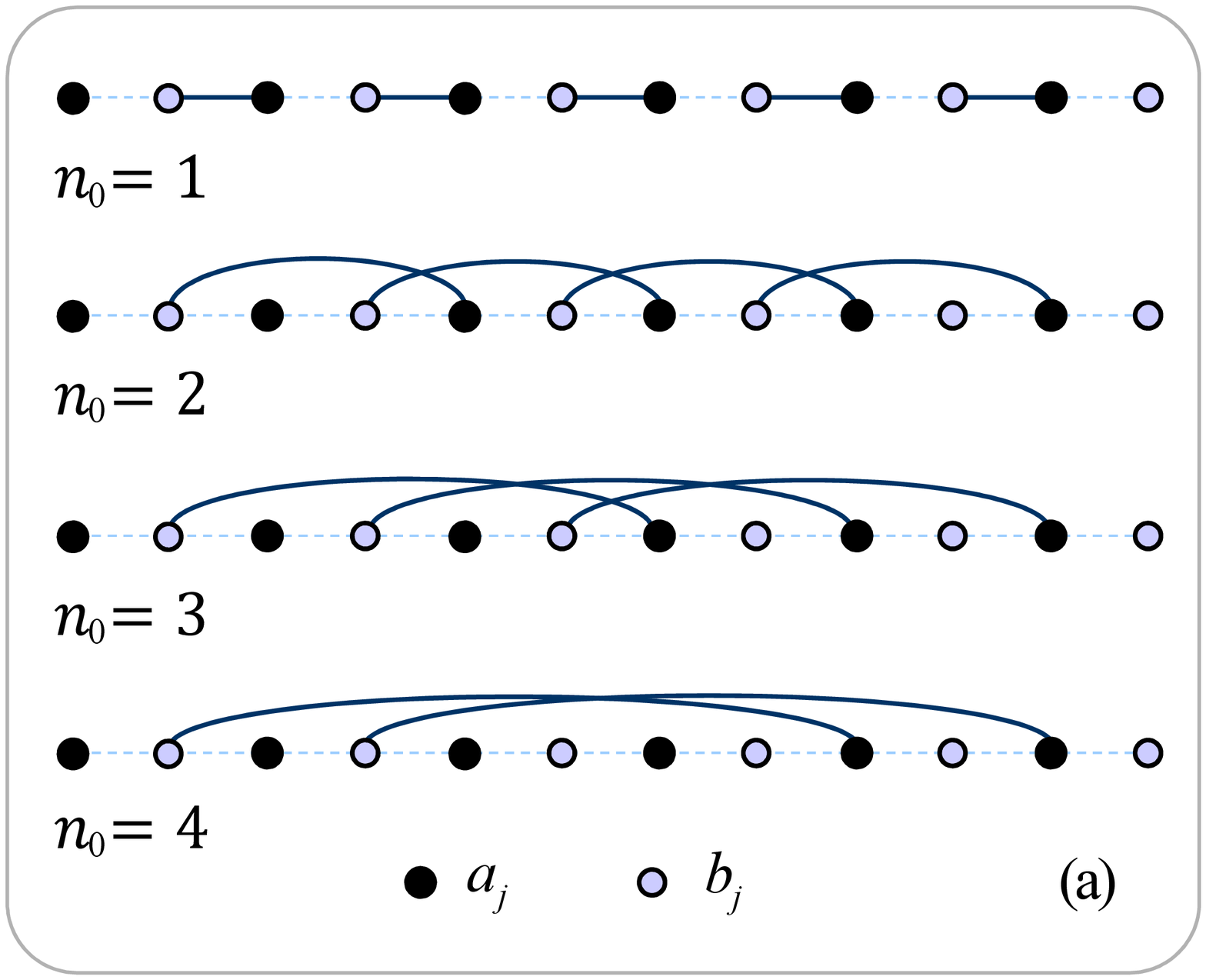} %
\includegraphics[ bb=20 355 575 810, width=0.45\textwidth, clip]{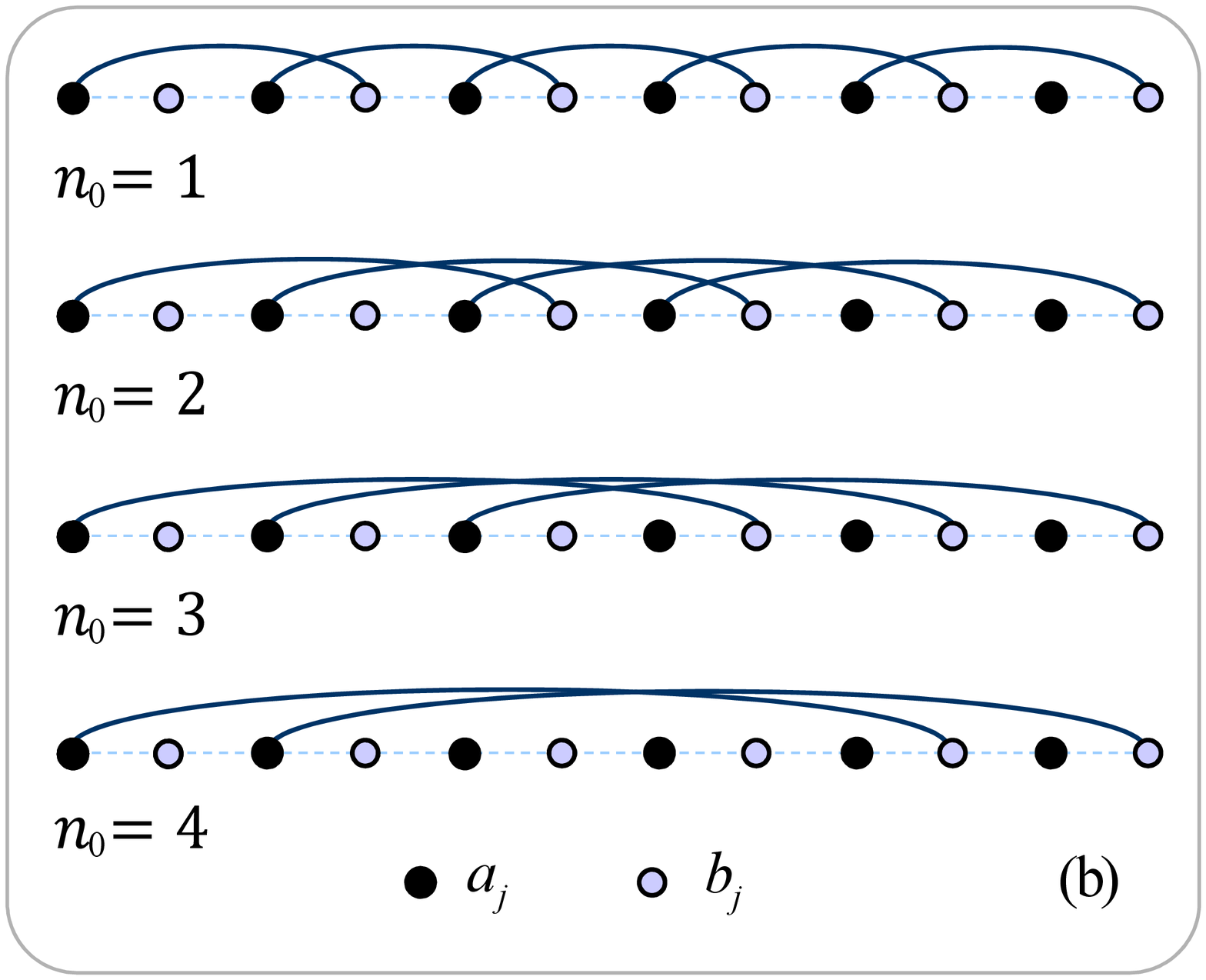}
\caption{(Color online) The structures of Majorana lattices for $h_{n_{0}}$\
defined in Eq. (\protect\ref{h_n}) with $n_{0}=1,2,3,4$. We find that $%
h_{n_{0}}$\ contains $2n_{0}$ isolated sites, allowing the existence of zero
modes.}
\label{fig2}
\end{figure*}

\begin{figure*}[tbp]
\includegraphics[ bb=13 53 530 484, width=0.45\textwidth, clip]{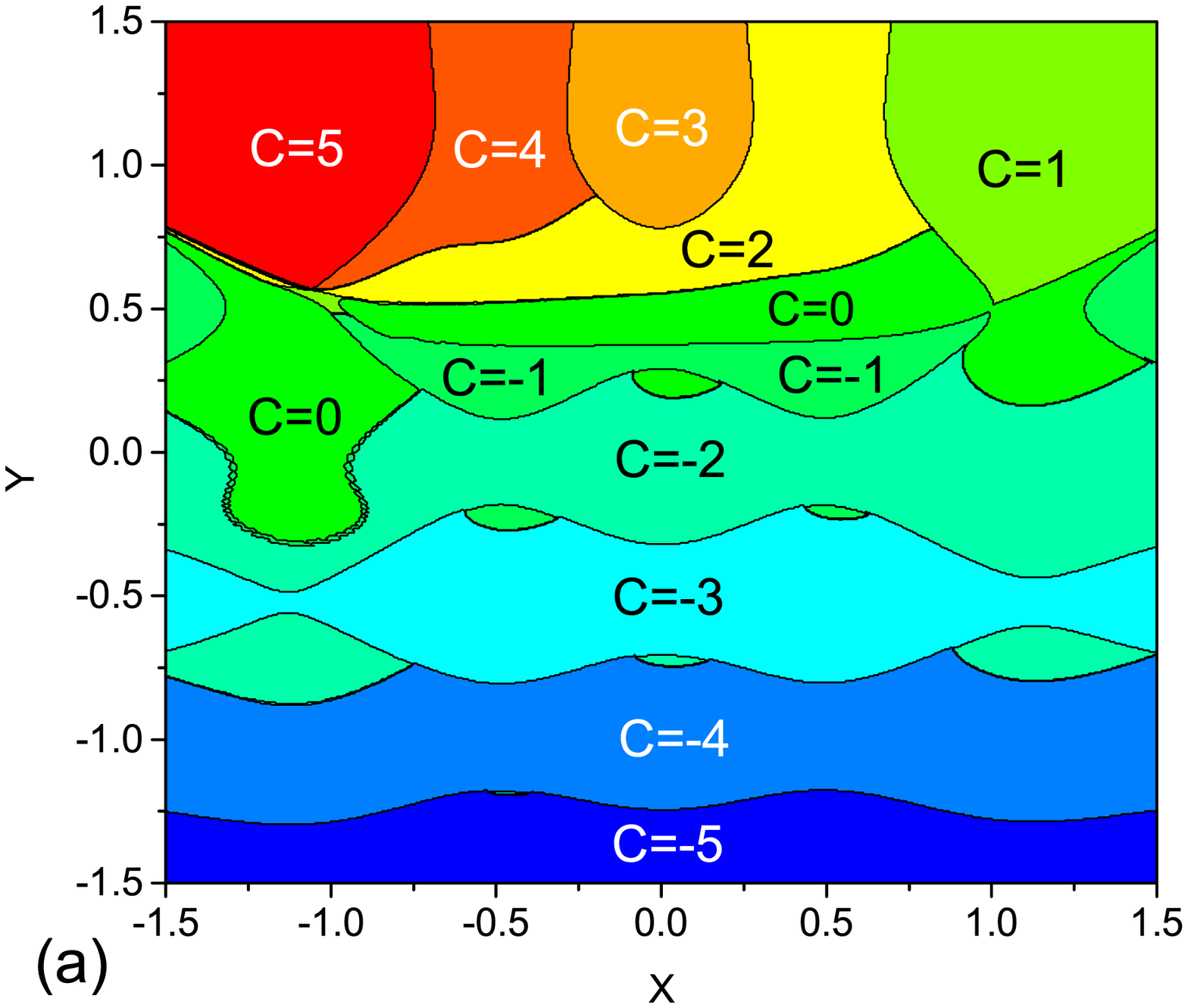} %
\includegraphics[ bb=13 53 530 484, width=0.45\textwidth, clip]{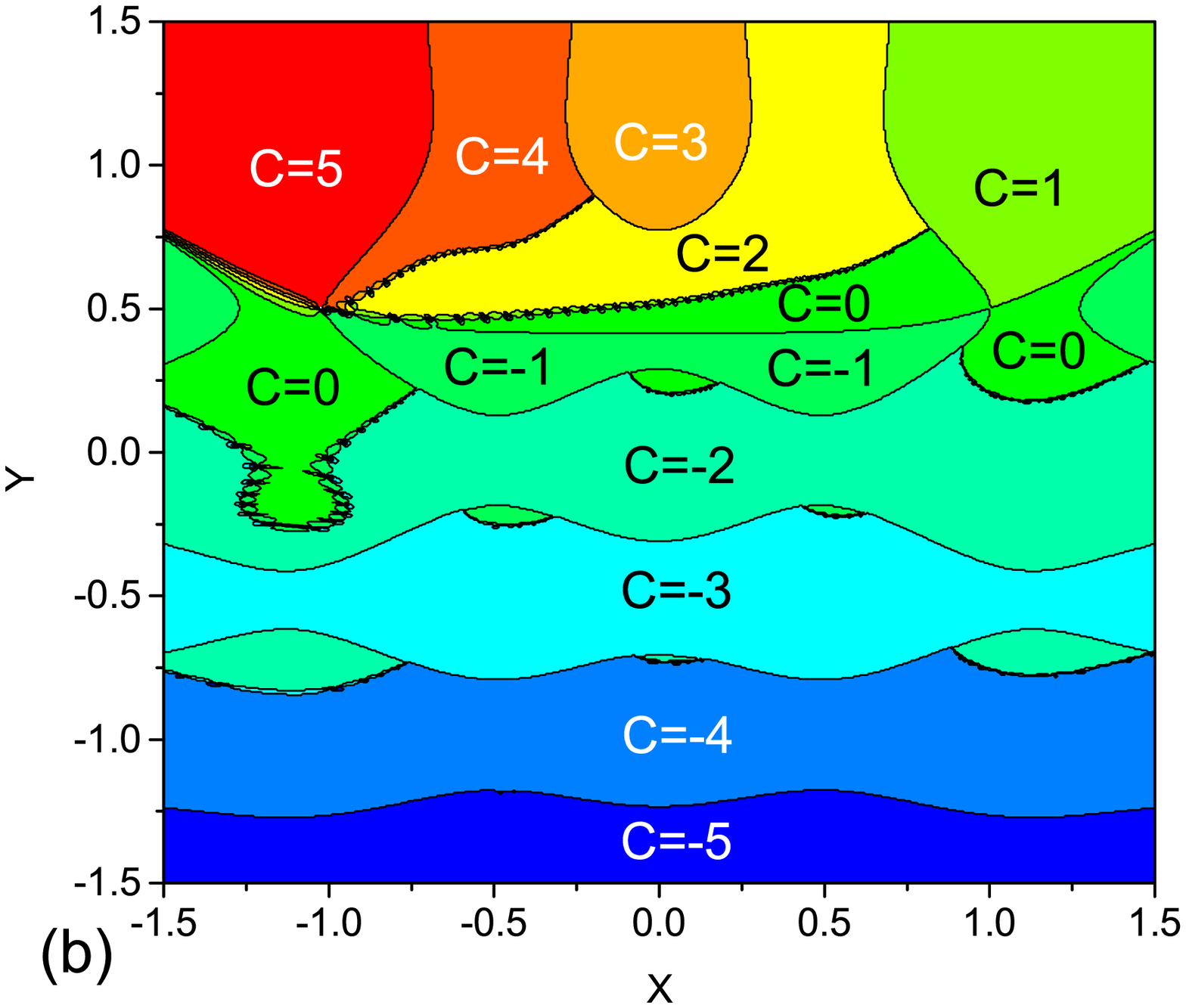}
\caption{(Color online) Phase diagrams for system with parameters satisfying
the equation (\protect\ref{chern_number}) identified by Chern numbers. The
Chern number is\ obtained by two ways: (a) It is computed by the winding
numbers from formula in Eq. (\protect\ref{winding number}). (b) It is
computed by the number of zero modes with the sign of Majorana charge
defined in Eq. (\protect\ref{charge}). The result is obtained by exact
diagonalization for Hamiltonian in Eq. (\protect\ref{h}) on $N=200$ chain.}
\label{fig3}
\end{figure*}

\section{Majorana charge of zero mode}

\label{Majorana charge of zero mode}The above results indicates that the
quantum phase of the model $H$ exhibits topological characterization.
Another way to unveil the hidden topology behind the model is exploring the
zero modes of the corresponding Majorana Hamiltonian. Consider the system
with open boundary conditions with the corresponding spinless fermion
representation $H_{\mathrm{ch}}$ in Eq. (\ref{Hch}).

We introduce Majorana fermion operators%
\begin{equation}
a_{j}=c_{j}^{\dagger }+c_{j},b_{j}=-i\left( c_{j}^{\dagger }-c_{j}\right) ,
\label{ab}
\end{equation}%
which satisfy the relations%
\begin{eqnarray}
\left\{ a_{j},a_{j^{\prime }}\right\} &=&2\delta _{j,j^{\prime }},\left\{
b_{j},b_{j^{\prime }}\right\} =2\delta _{j,j^{\prime }}, \\
\left\{ a_{j},b_{j^{\prime }}\right\} &=&0,a_{j}^{2}=b_{j}^{2}=1.
\end{eqnarray}%
The inverse transformation is
\begin{equation}
c_{j}^{\dagger }=\frac{1}{2}\left( a_{j}+ib_{j}\right) ,c_{j}=\frac{1}{2}%
\left( a_{j}-ib_{j}\right) .
\end{equation}%
Then the Majorana representation of the Hamiltonian is%
\begin{equation}
H=i\sum\limits_{n=1}^{M}\sum\limits_{j=1}^{N-n}\left(
J_{n}^{x}b_{j}a_{j+n}-J_{n}^{y}a_{j}b_{j+n}\right)
+ig\sum\limits_{j=1}^{N}a_{j}b_{j}.
\end{equation}%
We write down the Hamiltonian in the basis $\psi ^{T}=(a_{1},$ $b_{1},$ $%
a_{2},$ $b_{2},$ $a_{3},$ $b_{3},$ $...)$ and see that%
\begin{equation}
H=\psi ^{T}h\psi ,
\end{equation}%
where $h$\ represents a $2N\times 2N$ matrix. Here matrix $h$\ is explicitly
written as%
\begin{eqnarray}
h &=&\frac{i}{2}[\sum\limits_{n=1}^{M}\sum_{l=1}^{N-n}(J_{n}^{x}\left\vert
2l\right\rangle \left\langle 2\left( l+n\right) -1\right\vert \\
&&-J_{n}^{y}\left\vert 2l-1\right\rangle \left\langle 2\left( l+n\right)
\right\vert )+g\sum_{l=1}^{N}\left\vert 2l-1\right\rangle \left\langle
2l\right\vert +\mathrm{h.c.}],  \notag
\end{eqnarray}%
where basis $\left\{ \left\vert l\right\rangle ,l\in \left[ 1,2N\right]
\right\} \ $is an orthonormal complete set, $\langle l\left\vert l^{\prime
}\right\rangle =\delta _{ll^{\prime }}$. By taking a local unitary
transformation%
\begin{equation}
i\left\vert 2l\right\rangle \rightarrow \left\vert 2l\right\rangle
,\left\vert 2l-1\right\rangle \rightarrow \left\vert 2l-1\right\rangle ,
\end{equation}%
matrix $h$\ can be expressed as a simpler form with real matrix elements,%
\begin{eqnarray}
h &=&\frac{1}{2}[\sum\limits_{n=1}^{M}\sum_{l=1}^{N-n}(J_{n}^{x}\left\vert
2l\right\rangle \left\langle 2\left( l+n\right) -1\right\vert  \label{h} \\
&&+J_{n}^{y}\left\vert 2l-1\right\rangle \left\langle 2\left( l+n\right)
\right\vert )-g\sum_{l=1}^{N}\left\vert 2l-1\right\rangle \left\langle
2l\right\vert +\mathrm{h.c.}],  \notag
\end{eqnarray}%
which describes a tight-binding chain with long-range hopping.

We note that a winding number or Chern number has the positive or negative\
sign, denoting different phases, respectively. However, the number of zero
modes is always positive.\ It is expected to have another quantity, which
allows us to discern two phases with opposite Chern numbers, replacing the
number of zero modes. To this end, we introduce the concept of Majorana
charge for the first time to characterize the quantum phase. The magnitude
of a Majorana charge is defined by the distribution of particle probability
in zero-mode states, and the sign is defined by the type of Majorana
fermions, $a_{j}$\ or $b_{j}$. The exact expression of a Majorana charge is%
\begin{equation}
\mathcal{M}=\sum_{\alpha }\left\langle \alpha \right\vert (\widehat{M}_{+}-%
\widehat{M}_{-})\left\vert \alpha \right\rangle ,  \label{charge}
\end{equation}%
where $\left\vert \alpha \right\rangle $ denotes the zero mode state, and $%
\widehat{M}_{\pm }$ denotes the particle number operator of Majorana
fermion, which is defined as\
\begin{eqnarray}
\widehat{M}_{+} &=&\sum_{l=1}^{N_{+}}(\left\vert 2l-1\right\rangle
\left\langle 2l-1\right\vert \\
&&+\left\vert 2N+2-2l\right\rangle \left\langle 2N+2-2l\right\vert )  \notag
\\
\widehat{M}_{-} &=&\sum_{l=1}^{N_{-}}(\left\vert 2l\right\rangle
\left\langle 2l\right\vert \\
&&+\left\vert 2N+1-2l\right\rangle \left\langle 2N+1-2l\right\vert ),  \notag
\end{eqnarray}%
with%
\begin{equation}
N_{\pm }=\frac{N}{2}\pm \frac{1+\left( -1\right) ^{N+1}}{4}.
\end{equation}%
For even or large $N$ case, one can simply take $N_{\pm }=M$, since the
particles distribute mainly at two ends.

Now we focus on the relation between $\mathcal{M}$\ and Chern number. We
start our investigation from simple cases with $J_{n}^{y}=J_{n\neq
n_{0}}^{x}=0$ and $g=0$. The corresponding loop in the auxiliary space
reduces to%
\begin{equation}
\left\{
\begin{array}{c}
x_{n_{0}}\left( k\right) =J_{n_{0}}^{x}\cos \left( n_{0}k\right) \\
y_{n_{0}}\left( k\right) =J_{n_{0}}^{x}\sin \left( n_{0}k\right)%
\end{array}%
\right. ,
\end{equation}%
which is the superposition of $n_{0}$ identical circles with winding number $%
\mathcal{N=}$ $n_{0}$. The corresponding matrix in Majorana fermion
representation is%
\begin{equation}
h_{n_{0}}=\frac{J_{n_{0}}^{x}}{2}(\sum_{l=1}^{N-n_{0}}\left\vert
2l\right\rangle \left\langle 2\left( l+n_{0}\right) -1\right\vert +\mathrm{%
h.c.}).  \label{h_n}
\end{equation}%
Obviously, $h_{1}$\ describes a dimerized chain, which possesses two zero
modes and Majorana charge $\mathcal{M=}2$. Furthermore, it can be shown that
the spectrum of $h_{n_{0}}$\ possesses $2n_{0}$ zero modes. Based on the
above analysis, we know that when we consider other simple cases with $%
J_{n}^{x}=J_{n\neq n_{0}}^{y}=0$ and $g=0$. The corresponding loop in the
auxiliary space reduces to%
\begin{equation}
\left\{
\begin{array}{c}
x_{n_{0}}\left( k\right) =J_{n_{0}}^{y}\cos \left( n_{0}k\right) \\
y_{n_{0}}\left( k\right) =-J_{n_{0}}^{y}\sin \left( n_{0}k\right)%
\end{array}%
\right. ,
\end{equation}%
which has winding number $-n_{0}$. On the other hand, we have

\begin{equation}
\overline{h}_{n_{0}}=-\frac{J_{n_{0}}^{y}}{2}(\sum_{l=1}^{N-n_{0}}\left\vert
2l-1\right\rangle \left\langle 2\left( l+n_{0}\right) \right\vert +\mathrm{%
h.c.}),  \label{-h_n}
\end{equation}%
which can be shown that the spectrum of $\overline{h}_{n_{0}}$\ possesses $%
2n_{0}$ zero modes and the Majorana charge $\mathcal{M}=-2n_{0}$\textbf{.}

Straightforward derivations show that%
\begin{equation}
\lbrack \left\vert 2j-1\right\rangle \left\langle 2j-1\right\vert
,h_{n_{0}}]=0,
\end{equation}%
for $j<n_{0}+1$, and%
\begin{equation}
\lbrack \left\vert 2j\right\rangle \left\langle 2j\right\vert ,h_{n_{0}}]=0,
\end{equation}%
for $j>N-n_{0}$. This indicates that there are always $2n_{0}$\ isolated
sites in the ending region of the chain, resulting $2n_{0}$\ eigenstates
with zero energy. This analysis is applicable for $\overline{h}_{n_{0}}$. In
both situations, the Majorana charge equals to the winding number and Chern
number. In Fig. \ref{fig2}. the structures of Majorana\ lattices for $%
h_{n_{0}}$\ and $\overline{h}_{n_{0}}$\ with $n_{0}=1,2,3$, and $4$ are
schematically illustrated. We find that $h_{n_{0}}$\ and $\overline{h}%
_{n_{0}}$\ contain $2n_{0}$ isolated sites, allowing the existence of zero
modes. In the present stage, we cannot provide a proof for general case. We
explore the general case by exact numerical simulations. We calculate the
eigenvalues and Majorana charges for the systems with the parameters listed
in Table I (a)-(i) for finite $N$. Numerical results indicate that the
Majorana charges accord with the winding numbers or Chern numbers.

We demonstrate the richness of the phase diagram by a toy model with
parameters satisfying the equations%
\begin{eqnarray}
J_{n}^{x} &=&\exp \{-4[x-\frac{1}{2}\left( 3-n\right) ]^{2}\},  \label{J_x}
\\
J_{n}^{y} &=&2\left( y-1\right) \exp \{-4[y-\frac{1}{2}\left( 2-n\right)
]^{2}\}.  \label{J_y}
\end{eqnarray}%
The phase diagrams are obtained by Chern numbers for given $\left(
J_{n}^{x},J_{n}^{y}\right) $, which are computed in two different ways. On
the one hand, one can calculate the winding number through the numerically
integration in Eq. (\ref{winding number}), which has been shown to be equal
to Chern number. On the other hand, one can figure out the number of zero
modes by exact diagonalization of the Majorana matrix for finite $N$. The
sign of Chern number can be determined by the sign of the corresponding
Majorana charge defined in Eq. (\ref{charge}).

We have established our main results, and a few comments are in order.
First, notice that the zero mode states calculated here are not at exact
zero-energy for finite system except the cases for Hamiltonians in Eqs. (\ref%
{h_n}) and (\ref{-h_n}). Accordingly, the Majorana charges are not located
at exact edges. Second, we would like to point out that the sign of $M$\ is
not absolute, depending on the definition of $M$. If we take $\widehat{M}%
_{\pm }\rightarrow \widehat{M}_{\mp },$\ we will have $M\rightarrow -M$.
However, the relative sign of Majorana charge is meaningful: different signs
indicate different phases. Then when we say $c=N=M/2$, a suitable 2D-to-3D
extension and Majorana representation should be chosen.

\section{Summary}

\label{Summary}We have studied the topological characterization of QPTs in a
family of exactly solvable Ising models with short- and long-range
interactions. We calculate the Chern number and winding number for the
models with periodic boundary conditions. We have shown exactly that the
Chern number and winding number are identical and can be utilized to
characterize the phase diagram. This conclusion is applicable for more
generalized systems. We also calculate the Majorana mode charge analytically
and numerically. Our results indicate that the three numbers are equivalent.
Although our conclusion is obtained for specific models, it reveals the
possible connection between traditional and topological QPTs.

\end{document}